\documentclass[pre,aps,twocolumn]{revtex4-1}
\usepackage[latin1]{inputenc}
\usepackage[english]{babel}
\usepackage{graphicx}
\usepackage{color}
\usepackage{amsmath}
\usepackage{amssymb}

\begin{document}

\title{\bf Dissipative and stochastic geometric phase of a qubit within a
canonical Langevin framework}
\author{Pedro Bargue\~no $^{1,*}$ and Salvador Miret--Art\'es $^{2}$}
\affiliation{$^{1}$ Departamento de F\'{\i}sica de Materiales,
Universidad Complutense de Madrid, {\it 28040}, Madrid, Spain
($*$ p.bargueno@fis.ucm.es)
\\
$^{2}$ Instituto de F\'{\i}sica Fundamental, Consejo Superior de
Investigaciones Cient\'{\i}ficas, {\it 28006}, Madrid, Spain
}
\begin{abstract}
Dissipative and stochastic effects in the geometric phase of a
qubit are taken into account using a geometrical description of
the corresponding open--system dynamics within a canonical
Langevin framework based on a Caldeira--Leggett like Hamiltonian.
By extending the Hopf fibration $S^{3}\to S^{2}$ to include such
effects, the exact geometric phase for a dissipative qubit is
obtained, whereas numerical calculations are used to include
finite temperature effects on it.
\end{abstract}

\maketitle {\it Introduction}. The concept of geometric phase (GP)
in quantum systems was proposed by Berry \cite{Berry1984} when he
studied the dynamics of an isolated quantum system that undergoes
an adiabatic cyclic evolution. This cyclic evolution is due to the
variation of parameters of the Hamiltonian and is accompanied by a
change of the wavefunction by an additional phase factor which
depends only on the geometric structure of the space of
parameters. The underlying mathematical structure behind GPs was
pointed out almost simultaneously by Simon \cite{Simon1983}. Soon
after, the generalization to non--adiabatic cyclic evolution was
carried out by Aharonov and Anandan \cite{Aharonov1987} and
non-cyclic evolution and sequential projection measurements by
Samuel and Bhandary \cite{Samuel1988}. Although GPs have been
observed in the laboratory \cite{Tomita1986,Bitter1987,Leek2007},
realistic quantum systems are always subject to decoherence due to
their surroundings. Therefore, the extension of the GP to the case
of open quantum systems becomes fundamental. The first formal
extension of the GP was carried out by introducing the concept of
parallel transport along density operators \cite{Uhlmann1986}. In
a more physical context, the concept of GP was generalized for
non--degenerated mixed states \cite{Sjoqvist2000} and for
degenerate mixed--states under unitary evolution
\cite{Carollo2003}. Using a kinematic approach \cite{Mukunda1993},
GPs for mixed states in non--unitary evolution were addressed
\cite{Tong2004}. Within a spin--boson model, GPs in open quantum
systems have also been calculated \cite{Whitney2004a} together
with a study of their geometric nature \cite{Whitney2004b}. A
different approach was introduced in Ref. \cite{Marzlin2004},
where the GP was described by a distribution. In classical
physics, the counterpart of the Aharonov--Anandan (or Berry) phase
was early discovered by Hannay \cite{Hannay1985}. Regarding
classical dissipative systems, GP shifts have been defined in
dissipative limit cycle evolution
\cite{Kepler1991,Ning1992,Landsberg1992}, showing that they can be
identified with the classical Hannay angle in an extended phase
space \cite{Sinitsyn2008}. Moreover, GPs can be constructed for a
purely classical adiabatically slowly driven stochastic dynamics
\cite{Sinitsyn2007a,Sinitsyn2007b,Sinitsyn2009}.

In this Letter we tackle the problem of including dissipative and
stochastic effects in the GP of a qubit. Our study will be based
on a geometrical description for a non--isolated qubit within a
canonical Langevin framework (see \cite{Dorta2012} and references
therein) using a Caldeira--Leggett like Hamiltonian
\cite{Leggett1987}.

{\it Mathematical preliminaries}. It is well known  that if
$|\Psi\rangle$ represents a normalized $n$--level system, then
$|\Psi\rangle \in S^{2n-1}$. Thus, the geometry of
odd--dimensional spheres is related to the quantum mechanics of
finite--dimensional Hilbert spaces. In fact, the celebrated Hopf
fibration \cite{Hopf1931} relates the quantum and classical
description of qubits by means of the map $\pi: S^{3}\to S^{2}$.
This map can be understood as a composition, $\pi=\Xi  \circ
\Omega$, where $\Omega:S^{3}\subset \mathbb{C}^{2}\to
\mathbb{C}P^{1}$ sends an element of $\mathbb{C}^{2}$ to its
equivalence class and $\Xi:\mathbb{C}P^{1}\left(=\mathbb{C}\cup
\infty \right)\to S^{2}$ is given by the stereographic projection.
It can be shown that the Hopf map can be written in terms of the
Pauli matrices as $\pi\left(|\Psi\rangle \in S^{3}\right)=
\left(\langle \Psi|\hat\sigma_{x}|\Psi\rangle,\langle
\Psi|\hat\sigma_{y}|\Psi\rangle,\langle
\Psi|\hat\sigma_{z}|\Psi\rangle \right) \in S^{2}$, where $\langle
\Psi|\hat\sigma_{x}|\Psi\rangle^{2}+\langle
\Psi|\hat\sigma_{y}|\Psi\rangle^{2} +\langle
\Psi|\hat\sigma_{z}|\Psi\rangle^{2}=1$. Thus, from the Hopf map it
can be shown that quantum and classical mechanics may be embedded
in the same formulation. Specifically, for a qubit, the Strocchi
map \cite{Strocchi1966} is exactly the Hopf map previously
described. However, the $S^{7}\to S^{4}$ Hopf map, which is an
entanglement--sensitive fibration, does not have a classical
analog \cite{Bentssonbook}. As there is a map $S^{2n-1}\to
\mathbb{C}P^{n-1}$ and the complex projective space has a natural
symplectic structure ($\mathbb{C}P^{n-1}$ is a K\"ahler manifold),
$n$--state systems have a well-defined classical correspondence,
which is given by the Strocchi map \cite{Strocchi1966}. Thus, one
can derive a classical Hamiltonian function for a $n$--level
system by defining, for example, appropriate $n$ action--angle
coordinates in $\mathbb{C}P^{n-1}$ \cite{Oh1994}. For example, as
shown in \cite{Chruszinskybook}, the symplectic structure of
$S^{2}$ is responsible for the Aharonov--Anandan GP.

Using the pair of action--angle coordinates $(I,\Phi)$ on $S^{2}$,
the Hopf map can be expressed as $\pi\left(|\Psi\rangle\right)=
\left(\sqrt{1-I^{2}}\cos\Phi,\sqrt{1-I^{2}}\sin \Phi,I \right)$,
where $|\Psi\rangle=a_{1}|1\rangle +a_{2}|2\rangle$
($a_{j}=|a_{j}|e^{i\phi_{j}}\in \mathbb{C}$), $I\equiv
|a_{1}|^{2}-|a_{2}|^{2}$ and $\Phi\equiv \phi_{1}-\phi_{2}$. Thus,
the Hamiltonian operator $\hat H= \sum_{i}\eta_{i}\hat
\sigma_{i}$, where $\hat\sigma_{i}$ are the Pauli matrices and
$\eta_{i}\in \mathbb{R}$, can be Hopf--mapped to a Hamiltonian
function, $H_{0}$, given by $H_{0}=
-2\sqrt{1-I^{2}}\left(\eta_{1}\cos\Phi +\eta_{2}\sin\Phi\right) +
2\eta_{3}I$ (this is the Meyer--Miller-Stock-Thoss Hamiltonian
\cite{Meyer1979,Stock1997}, widely used in molecular physics).

{\it Dissipative--stochastic Hopf fibration of $S^{3}$}.  Our
study will be based on a Caldeira--Leggett \cite{Leggett1987} like
Hamiltonian for a qubit in the Langevin framework (see
\cite{Dorta2012} and references therein), which can be expressed
as
\begin{equation}
H = H_{0}+
\frac{1}{2}\sum_{i}\left(p^{2}_{i}+x^{2}_{i}\omega_{i}^{2}\right)
-\Phi\sum_{i}c_{i}x_{i}+\sum_{i}\Phi^{2}c^{2}_{i} \label{hcal}
\end{equation}
where the oscillator mass has been taken to be one and $c_{i}$ are
the system--bath coupling constants. This model takes into account
a renormalization term due to the interaction with the
environment.

Let us start with pure and normalized quantum states. Using the
$(I,\Phi)$ action--angle coordinates, it is easy to see that the
unit-radius sphere $S^{2}$ is the set of  points satisfying
$\left(-\sqrt{1-I^{2}}\cos\Phi\right)^{2}+\left(-\sqrt{1-I^{2}}\sin
\Phi\right)^{2}+I^{2}=\nolinebreak 1$. The radius of this sphere
remains constant along the time due to energy conservation. That
is, the corresponding Hamiltonian function, given by
$H_{0}=-\sqrt{1-I^{2}}\left(\cos\Phi+\sin \Phi\right)+I$ remains
constant in time (we have taken $\eta_{i}=1$ for simplicity). When
Ohmic dissipation is assumed, and after removing the bath
variables, the corresponding equations of motion issued from Eq.
(\ref{hcal}) are
\begin{eqnarray}
\label{stochohm1} \dot I &=&-\sqrt{1-I^2}(\sin \Phi+\cos\Phi) - 2\gamma\, \dot \Phi(t) +\xi(t)\nonumber \\
\dot \Phi &=&\frac{I}{\sqrt{1-I^2}}(\cos \Phi+\sin\Phi)+ 1,
\end{eqnarray}
which correspond to the following effective Hamiltonian function
$H(t)=-\sqrt{1-I^{2}}\left(\cos\Phi+\sin \Phi\right)+I+2\gamma\Phi
\dot \Phi -\xi(t) \Phi$. This stochastic dynamics can be
interpreted in terms of a stochastic Bloch sphere
$S^{2}(\gamma,\xi)$ defined by

\begin{figure}[h!]
\begin{center}
\includegraphics*[height=6cm,width=6cm]{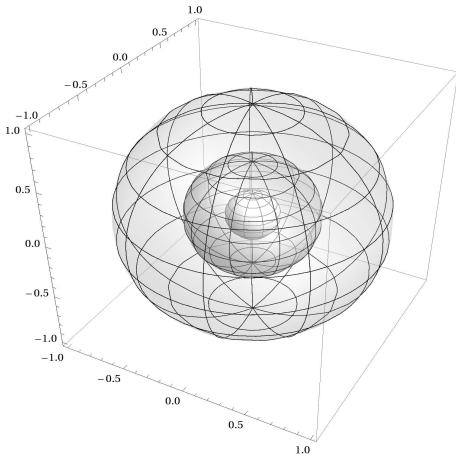}
\end{center}
\caption{\label{fig1}Representation of the dissipative--induced
time dependence of the Bloch sphere for $\xi=0$. Time is running
in the radial direction.}
\end{figure}

\begin{widetext}
\begin{equation}
\left(\sqrt{1-I^{2}}\cos\Phi\right)^{2}+\left(\sqrt{1-I^{2}}\sin
\Phi\right)^{2}+I^{2}=1-\gamma\frac{d}{dt}(\Phi^{2}) +\xi(t)\Phi ,
\label{general}
\end{equation}
\end{widetext}

\noindent where $\gamma$ is the friction constant, $\xi(t)$ is a
stochastic Gaussian process. The time--dependent radius is given
by $R_t^{2}(\gamma,\xi)=1-\gamma\frac{d}{dt}(\Phi^{2})+\xi\Phi$.
Thus, dissipative and stochastic effects make the Bloch sphere
breathe, Eq. (\ref{general}), by changing its radius in time. This
radius is bounded for $\xi=0$, $R_t(\gamma,0) \le 1$. The equality
is reached at $t=0$ and at asymptotic times, where thermal
equilibrium is reached (in this case, a point in the unit radius
sphere represents a pure state). Otherwise, mixed states are
represented at each instant of time as points in different spheres
of variable radius, as shown in Fig. (\ref{fig1}).
When $\xi\ne0$, the square radius becomes a stochastic variable
which could reach $R_t(\gamma,\xi)\ge1$ for a particular
phase--space trajectory.

The breathing of the Bloch sphere can be geometrized by extending
its round metric by adding both dissipation and noise. If these
terms are included, the metric of $S^{2}(\gamma,\xi)$ can be
written as $ds^{2}\left[S^{2}(\gamma,\xi)\right]=
R_t^2(\gamma,\xi) ds^{2}\left[S^{2}\right]$, where
$ds^{2}\left[S^{2}\right]=\frac{dI^{2}}{1-I^{2}}+(1-I^{2})d\Phi^{2}$
is the round--metric for $S^{2}$ in action--angle coordinates.


In order to calculate the stochastic GP, let us start with pure
states. If we choose an orthonormal moving frame field
$\sigma^{1}=\frac{dI}{\sqrt{1-I^{2}}}$,
$\sigma^{2}=\sqrt{1-I^{2}}d\Phi$, then
$d\sigma^{2}=-\frac{I}{\sqrt{1-I^{2}}}dI\wedge
d\Phi=-I\sigma^{1}\wedge d\Phi=I d\Phi\wedge\sigma^{1}$. Using
Cartan's first structure equation, we obtain the only
non-vanishing connection one--form, $\omega=I d\Phi$. Thus, the
dynamic phase is obtained as $\phi_{d}=\oint \omega = \oint I
d\Phi = \pi I = \pi \cos \theta$. If $S^{2}$ is considered as an
embedded submanifold of $S^{3}$, taking into account that the spin
connection of $S^{3}$ is $\tilde \omega=d\Xi+I d\Phi$, where $\Xi$
is the extra Euler angle which parametrizes the third dimension,
it can be shown \cite{Bentssonbook} that $S^{2}$ inherits from
$S^{3}$ the connection one--form $\tilde \omega=(I- 1)d\Phi$. Thus,
the GP is expressed as $\phi_{g}=\oint \tilde \omega$. Now, if
dissipation and noise are taken into account, the orthonormal
moving frame field is given by $R_t(\gamma,\xi)\sigma^{i}$ with
$i=1,2$. In this case, the only non-vanishing connection one--form
is $\omega(\gamma,\xi)=R_t(\gamma,\xi)I
d\Phi=R_t(\gamma,\xi)\omega$. This allows us to define the
stochastic dynamic phase acquired after a cycle of period $T$ as
$\phi_{d}(\gamma,\xi)=\oint \omega (\gamma,\xi)=\oint
R_t(\gamma,\xi)I(T) d\Phi=
\cos\theta(T)\oint\sqrt{1-\gamma\frac{d}{dt}(\Phi^{2})+\xi
\Phi}d\Phi$. Therefore, the stochastic connection one--form, which
can be defined as $\tilde \omega (\gamma,\xi)=\left[I
R_t(\gamma,\xi)- 1\right] d\Phi$, leads to the corresponding GP

\begin{widetext}
\begin{equation}
\label{eqdef}
\phi_{g}(\gamma,\xi) =\oint \tilde \omega (\gamma,\xi)=\oint
\left[I(T) R_t(\gamma,\xi)- 1\right] d\Phi =\oint\left(
I(T)\sqrt{1-\gamma\frac{d}{dt}(\Phi^{2})+\xi \Phi}- 1\right)d\Phi
.
\end{equation}
\end{widetext}

\noindent Notice that the dynamics driven by this kind of
Caldeira--Leggett like coupling could be interpreted in terms of a
mapping between conformal spheres with conformal unitary group
fibers at each instant. Moreover, the GP defined by Eq.
(\ref{eqdef}) becomes $\tilde \phi_{g}(\gamma,\xi)=\oint\left(
I\sqrt{1-\gamma\frac{d}{dt}(\tilde\Phi^{2})+\xi \tilde\Phi}-
1\right) d\tilde\Phi$ under the transformation
$|\tilde\Psi\rangle=e^{i\alpha}|\Psi\rangle$ (or $\tilde
\Phi=\Phi+\alpha$). Thus, $\phi_{g}(\gamma,\xi)$ is a gauge
invariant quantity.

{\it Dissipative qubit (zero temperature)}. Let us consider a
simple qubit which can be represented by the Hamiltonian operator
$\hat H= \epsilon \hat \sigma_{z}$. The corresponding dissipative
dimensionless Hamiltonian function ($t\rightarrow 2\epsilon t$),
which can be written as $H(t)=I+\frac{\gamma}{2\epsilon}\Phi \dot
\Phi$, leads to the pair of coupled equations $\dot I = -
\frac{\gamma}{2\epsilon} \dot \Phi $ and $\dot \Phi = 1$ with
solutions $I(t)=I_{0}-\frac{\gamma}{2\epsilon}t$ and
$\Phi(t)=\Phi_{0} + t$, where $I_0$ and $\Phi_0$ are the initial
conditions of the action--angle variables. After a cycle of period
$T=2\pi$ (the scaled Rabi frequency is $\omega_{0}=1$), the
dissipative GP can be computed as

\begin{eqnarray}
\label{eqfin}
\phi_{g}(\gamma)&=&\oint \left[R_t(\gamma)I -1\right]d\Phi \nonumber \\
&=&\oint \left[I(T)\sqrt{1-\frac{\gamma}{2\epsilon}\Phi \dot \Phi}-1\right]d\Phi  \nonumber \\
&=&-\pi+\frac{4}{3}\left[\left(1-\frac{\pi\gamma}{2\epsilon}\right)^{3/2}-1\right]\left(\pi-\frac{\epsilon}{\gamma}
\cos\theta_{0}\right) , \nonumber \\
\end{eqnarray}

\noindent where $I_0 = \cos \theta_0$. Notice that the
non--dissipative GP is recovered when $\gamma\to 0$.

It is also interesting to remark that Eq. (\ref{eqfin}) is only
valid for $\gamma \pi/2\epsilon\le 1$, which reflects nothing but
the Caldeira--Leggett energy renormalization in an Ohmic
environment. Moreover, this energy renormalization can be
re--interpreted in terms of a Bloch sphere whose radius developes
harmonic oscillations. In order to show this statement in simple
terms, let us introduce the damping factor by means of the
phenomenological Caldirola--Kanai Hamiltonian
\cite{Razavi2005,Caldirola1941,Kanai1948} for a harmonic
oscillator (which is equivalent to the corresponding
Caldeira--Leggett Hamiltonian for the oscillator at zero
temperature). The Hamiltonian reads $H=(1/2) p^{2}
e^{-\frac{\gamma}{2\epsilon}\pi t}+ q^{2}
e^{\frac{\gamma}{2\epsilon}\pi t}$ (the $\pi$ factor has been
introduced to note that time evolution is cyclic). It is
straightforward to derive the corresponding equations of motion,
leading  to $\ddot
y+\left[1-\left(\frac{\gamma\pi}{2\epsilon}\right)^{2}\right]y=0$,
where $y=p,q$. Thus, the renormalized frequency due to the damping
term is
$\bar\omega^2=\omega_{0}^2-\left(\frac{\gamma\pi}{2\epsilon}\right)^{2}$
which has physical sense only when $\gamma \pi/2\epsilon\le 1$, as
Eq. (\ref{eqfin}) shows. Therefore, the radius of the Bloch sphere
developes harmonic oscillations with a renormalized frequency
(another interpretation is that $|I(t)|\le 1$ requires $\gamma
\pi/2\epsilon\le 1$).

The weak coupling limit of the GP given by Eq. (\ref{eqfin}) can be
expressed as
\begin{equation}
\phi_{g}(\gamma)= -\pi\left(1-\cos
\theta_{0}\right)-\frac{\gamma}{\epsilon}\left(\frac{\pi}{2}\right)^{2}\left(\cos
\theta_{0}+4\right) + \mathcal{O}(\gamma^{2}) .
\end{equation}
\noindent A direct comparison with other authors
\cite{Carollo2003,Tong2004,Marzlin2004} is not pertinent since
different models of dissipation and other approaches to calculate
the GP were used.

On the other hand, for this dissipative dynamics, information on
interference experiments can be straightforwardly extracted from
the probability density itself. The typical interference
intensity, $J_t$, evolves in time according to $J_t\propto
|\Psi(t)|^{2}\propto 1 + \sqrt{1-I^{2}(t)}\cos
\Phi(t)=1+\sqrt{1-\left(I_{0}-\frac{\gamma}{2\epsilon}t\right)^{2}}
\cos(t+\Phi_{0})$, depending critically on the ratio $\gamma / 2
\epsilon$ which is directly related to the GP given by Eq.
(\ref{eqfin}).

{\it Stochastic qubit (non--zero temperature)}. Noise effects can
be included in a simple way by assuming a Gaussian stochastic
process with distribution function $\rho (\xi;\beta) =
\sqrt{\frac{\beta}{2\pi }} \exp{\left(-\frac{\beta \xi^{2}}{2}\right)}$, where $\beta$ is the inverse of the temperature,
which is given in units of $2\epsilon$. Thus, for a qubit, the
squared radius is also a stochastic process given by
$R_t^2(\gamma,\xi)=1+(\xi-\frac{\gamma}{2\epsilon})\Phi$.
Therefore, the stochastic GP acquired after a cycle of period $T$
is given by

\begin{widetext}
\begin{equation}
\phi_{g}(\gamma;\beta)=\left( \cos\theta(T)-\frac{\gamma \pi}{2\epsilon}\right)\left[
\oint \int_{-\infty}^{\infty}\sqrt{1-\Phi\left(\frac{\gamma}{2\epsilon}-\xi\right)}\sqrt{\frac{\beta}{2\pi}}
\exp{\left(-\frac{\beta\xi^{2}}{2}\right)} d\xi d\phi
\right]-\pi.
\end{equation}
\end{widetext}

\noindent The previous integral in the noise variable has no
analytic solution. Thus, we firstly integrate in $\Phi$ obtaining

\begin{widetext}
\begin{equation}
\phi_{g}(\gamma;\beta)=\left( \cos\theta(T)-\frac{\gamma \pi}{2\epsilon}\right)\left[
\int_{-\infty}^{\infty}-\frac{2}{3}\left(\frac{\gamma}{2\epsilon}-\xi\right)^{-1}
\left[\left(1-\bigl\{\frac{\gamma}{2\epsilon}-\xi\bigl\}\pi\right)^{3/2}-1\right]
\sqrt{\frac{\beta}{2\pi}}
\exp{\left(-\frac{\beta\xi^{2}}{2}\right)} d\xi
\right]-\pi.
\end{equation}
\end{widetext}

\noindent To illustrate the computation of this integral (which
does not have any analytical solution either), only the GP for a
non--dissipative qubit at finite temperature will be considered.
Moreover, as the noise term becomes more important for high
temperatures, we assume that the mean of the Gaussian process is
$\beta^{-1}$. Therefore, the GP can be factorized as

\begin{figure}[h!]
\begin{center}
\includegraphics*[angle=-90,origin=c,height=7cm,width=8cm]{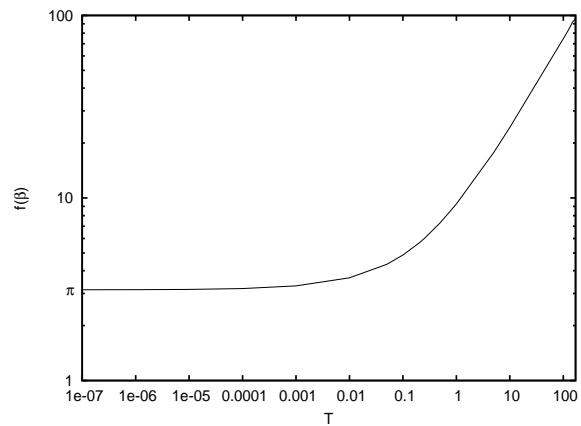}
\end{center}
\caption{\label{fig2} Thermal behavior of the GP obtained by
numerical integration of Eq. (\ref{eqtemp}) (double logarithmic
scale is used). Notice that $f(\beta)\to \pi$ at very low
temperatures.}
\end{figure}

\begin{widetext}
\begin{equation}
\label{eqtemp}
\phi_{g}(\beta)=\cos\theta_{0}\left[
\int_{-\infty}^{\infty}\frac{2}{3\xi}
\left[\left(1+\xi\pi \right)^{3/2}-1\right]
\sqrt{\frac{\beta}{2\pi}}
\exp{\bigl\{-\frac{\beta(\xi-\beta^{-1})^{2}}{2}\bigl\}} d\xi
\right]-\pi \equiv \cos \theta_{0} f(\beta)-\pi,
\end{equation}
\end{widetext}

\noindent where $f(\beta)$ encodes the thermal information of the
GP acquired by an isolated qubit. This function is depicted in
Fig. (\ref{fig2}) within a large range of temperatures, displaying
a linear behavior at high temperatures and a crossover at a
critical temperature given by $T_{c}\approx 2\epsilon$
($T_{c}\approx 1$ if adimensional temperatures are used). This
temperature corresponds to the energy difference of the two levels
of the qubit. Moreover, as $f(\beta)\to \pi$ at very low
temperatures, the GP acquired for the pure state at zero
temperature is recovered. Finally, notice that this type of
calculations can be straightforwardly extended to any dissipation
value.

In summary, a geometrical description of the dissipative and
stochastic  dynamics of a qubit within a Langevin formalism (with
a Caldeira--Leggett like coupling) has been developed. The Hopf
fibration $S^{3}\to S^{2}$ has been extended to include both
stochastic and dissipative effects in terms of a Bloch sphere
which develops harmonic oscillations (in the dissipative case).
This procedure has allowed us to define a gauge invariant
geometric phase, which has been computed for both dissipative and
stochastic cases.

This work has been funded by the MICINN (Spain) through Grant Nos.
CTQ2008--02578 and FIS2011--29596-C02-01. P. B. acknowledges a
Juan de la Cierva fellowship from the MICINN.

\end{document}